\documentclass[draft]{agujournal2019}
\usepackage{url} 
\usepackage{lineno}
\usepackage{soul}
\usepackage{pdflscape}	
\linenumbers
\draftfalse

\journalname{JGR: Space Physics}

\begin{document}
\nolinenumbers

%
%


\title{Evolution of the Earth's polar outflow from mid-Archean to present}

%
%




\authors{K.~G.~Kislyakova\affil{1}\thanks{T\"urkenschanzstrasse 17, 1180 Vienna, Austria}, C.~P.~Johnstone\affil{1}, M.~Scherf\affil{2}, M.~Holmstr\"om\affil{3}, I.~I.~Alexeev\affil{4}, H.~Lammer\affil{2}, M.~L.~Khodachenko\affil{2,5,6}, M.~G\"udel\affil{1}}

 \affiliation{1}{Department of Astrophysics, University of Vienna, Vienna, Austria}
 \affiliation{2}{Space Research Institute, Austrian Academy of Sciences, Graz, Austria}
 \affiliation{3}{Swedish Institute of Space Physics, Kiruna, Sweden}
 \affiliation{4}{Skobeltsyn Institute of Nuclear Physics, Moscow State University, Moscow, Russia}
 \affiliation{5}{Institute of Astronomy of the Russian Academy of Sciences, 119017, Moscow, Russia}
 \affiliation{6}{Institute of Laser Physics, SB RAS, Novosibirsk, 630090, Russia}





\correspondingauthor{K.~G.~Kislyakova}{kristina.kislyakova@univie.ac.at}




\begin{keypoints} 
\item Polar outflow escape of the nitrogen ions three gigayears ago increases by two orders of magnitude compared to its present value.
\item Polar outflow of oxygen ions from the Earth's open field line bundle varies greatly depending on the oxygen mixing ratio.
\item Polar outflow is governed primarily by the evolution of the solar short-wavelength radiation and the atmosphere's composition. 
\end{keypoints}

%
%

%
%


\begin{abstract}

The development of habitable conditions on Earth is tightly connected to the evolution of its atmosphere which is strongly influenced by atmospheric escape. We investigate the evolution of the polar ion outflow from the open field line bundle which is the dominant escape mechanism for the modern Earth. We perform Direct Simulation Monte Carlo (DSMC) simulations and estimate the upper limits on escape rates from the Earth's open field line bundle starting from three gigayears ago (Ga) to present assuming the present-day composition of the atmosphere. We perform two additional simulations with lower mixing ratios of oxygen of 1\% and 15\% to account for the conditions shortly after the Great Oxydation Event (GOE).
  
We estimate the maximum loss rates due to polar outflow three gigayears ago of $3.3 \times10^{27}$~s$^{-1}$ and $2.4 \times 10^{27}$~s$^{-1}$ for oxygen and nitrogen, respectively. The total integrated mass loss equals to 39\% and 10\% of the modern atmosphere's mass, for oxygen and nitrogen, respectively. According to our results, the main factors that governed the polar outflow in the considered time period are the evolution of the XUV radiation of the Sun and the atmosphere's composition. The evolution of the Earth's magnetic field plays a less important role. We conclude that although the atmosphere with the present-day composition can survive the escape due to polar outflow, a higher level of CO$_2$ between 3.0 and 2.0~Ga is likely necessary to reduce the escape.

\end{abstract}


%
%

%


%
%
%
%

\section{Introduction}

The maintenance of clement climate conditions on the Earth's surface through its history was facilitated by its nitrogen-dominated atmosphere. The mass and composition of a secondary (which is dominated by gases other than hydrogen) atmosphere depend on the balance of the outgassing from interiors, surface interaction and escape to space \cite{VanHoolst19}. Atmospheric escape from Earth, Venus, and Mars through the history of these planets was governed by two factors: first, the evolution of the solar short wavelength radiation and wind and second, the evolution of the atmosphere's composition, as some gases are more prone to escape than others (e.g., \citeNP{Lammer18,Gronoff20} and references therein). The Sun and its planets evolved together, but the outcome of this evolution was very different for the three planets. While Earth evolved to have an ocean and an atmosphere with a moderate pressure that is composed mostly of nitrogen, Mars and Venus lost most of their water and evolved into arid inhospitable worlds with the atmospheres dominated by CO$_2$. 

In the solar system, the current loss rates from Venus, Earth, and Mars are similar, but the dominant escape mechanisms differ. In all three cases, the escape is dominated by a non-thermal mechanism. On Venus, the dominant escape mechanism is ion pick-up, on Mars it is the outflow of hot atoms, and on Earth it is polar outflow from the open field line bundle (e.g., \citeNP{Barabash07,Amerstorfer17,Lammer18}). The Earth is unique among the three planets since it has an internal magnetic field that governs the processes in the magnetosphere and leads to the formation of the open field line bundles near the magnetic poles. 

Magnetic fields of rocky planets are believed to play an important role for planetary habitability \cite{McIntyre19}. On Earth, the ions mostly escape from the open field line bundles, which are absent for modern Mars and Venus since they do not possess intrinsic magnetic fields. In those cases, escape takes place through the plasma wake and from a boundary layer of the induced magnetosphere \cite{Barabash07,Lundin11,Haaland15,Gunell18}. The extent to which a magnetosphere protects its planetary atmosphere from the stellar wind erosion depends on how well it prevents energy and momentum exchange with the atmosphere and traps otherwise escaping plasma \cite{Blackman18}. It is not clear yet how important the magnetosphere is for protection of the planetary atmosphere (see discussion in \citeNP{Blackman18,Gunell18,Egan19}).

The intrinsic magnetic field of the Earth was present already very early in its history \cite{Tarduno10, Tarduno15}. Therefore, the evolution of the terrestrial atmosphere was coupled to the evolution of the Earth's magnetic field. Although the magnetic field was present as early as 4.0~Ga, it was weaker in the past \cite{Tarduno15}. The combined effect of the early Earth's weaker magnetic field and a stronger solar wind \cite{Johnstone15b,OFionnagain18} led to higher compression of the magnetosphere and to wider opening angles of the polar ovals than at present time \cite{Airapetian16} and both factors can enhance the escape rates. Since the atmosphere in the past was likely expanded due to enhanced solar XUV fluxes, one can expect even higher past escape rates. 

The Sun's X-ray and extreme ultraviolet radiation (together XUV; in this article defined as the entire range from X-rays to UV of 1--400~nm following \citeNP{Johnstone18}) heats up upper atmospheres, leads to their expansion and drives thermal escape \cite{Tian08,Lichtenegger10}. Multiple pathways of the early Sun's XUV and solar wind evolution were possible \cite{Tu15,Johnstone15b}. The amount of short-wavelength radiation a star produces is tightly linked to its rotational evolution, with stars born as fast rotators emitting more XUV radiation and likely having stronger winds than stars born as slow rotators. Although the investigation of the Sun's early history is far from being complete, there exist some indications that the Sun was born as a slow rotator \cite{Lammer20}.

Atmospheres with low CO$_2$ mixing ratios such as the atmosphere of the modern Earth are especially sensitive to high XUV fluxes since they lack efficient coolants in the mesosphere and thermosphere. This increases the temperature in the upper part of the atmosphere and enhances both thermal and non-thermal escape \cite{Kulikov07,Lichtenegger10,Johnstone19}. The effect of the XUV radiation on thermal atmospheric escape was studied in earlier works (e.g., \citeNP{Tian08,Johnstone18,Johnstone19}), but the combined effect of evolving atmospheric temperature and density profiles and the magnetic field was not studied in detail. Recently, \citeNP{Johnstone19} showed that the present-day Earth atmosphere would rapidly escape hydrodynamically if exposed to a very strong XUV flux of the Sun 4.4~Ga. In later times, such as three Ga and later, the thermal Jeans escape rate is much lower, so that a nitrogen-dominated atmosphere can potentially survive the thermal losses \cite{Johnstone18}, but the non-thermal losses in this time period were not investigated in detail. 

In this article, we study the evolution of the polar ion outflow rates from mid-Archean to present and estimate the maximum possible escape rates of oxygen and nitrogen starting from three Ga and up to the present time. We consider the effects of the evolving solar XUV output on the Earth's thermosphere and the resulting influences of these effects on the interactions between the atmosphere and the solar wind. We also consider the changes in the solar wind properties and the Earth's intrinsic magnetic field during this time. In Section~\ref{sec_methods}, we present our model for the upper atmosphere, exosphere, and magnetosphere, and describe the method we use to estimate the escape rates. In Section~\ref{sec_results}, we present our results and in Section~\ref{sec_conslusions}, we summarize our conclusions.


\section{Methods}
\label{sec_methods}

\subsection{Upper atmosphere model}

To calculate atmospheric losses in the exosphere and magnetosphere, it is necessary to know the properties of the planet's thermosphere since this determines the lower boundary conditions in our DSMC simulations of the exosphere.
For this purpose, we use thermospheric profiles calculated for the Earth exposed to the XUV radiation of the young Sun from the mid-Archean to present \cite{Johnstone18}. These profiles were calculated using the Kompot code, which is a first-principles model that calculates the thermal, chemical, and hydrodynamic structures of a planetary upper atmosphere for arbitrary planetary parameters, atmospheric compositions, and stellar inputs. The code includes heating from the absorption of stellar X-ray, UV, and IR radiation, heating from exothermic chemical reactions, electron heating from collisions with non-thermal photoelectrons, Joule heating, cooling from IR emission by several species, thermal conduction, and energy exchanges between the neutral, ion, and electron gases. The chemistry includes $\sim$500 chemical reactions, including 56 photoreactions, eddy and molecular diffusion, and advection. The code is able to reproduce the thermal and chemical structures of the atmospheres of Earth and Venus, and is a powerful tool to calculate the structure, ionization rate, and electrical conductivities of planetary ionospheres and is the most up-to-date model in the field. The Kompot code can be used to calculate hydrodynamic outflow from atmospheres \cite{Johnstone19,Johnstone20}, but this is not relevant for the Earth for the times that we study in this paper as the activity and the XUV flux of the Sun have already significantly declined by the mid-Archean \cite{Ribas05,Guedel07,Tu15}.  In general, the early evolutionary tracks for rotation and XUV radiation of the Sun is not relevant for the present article as all evolutionary tracks for Sun-like stars converge after the star reaches the age of approximately one gigayear. Therefore, we consider only one evolutionary track for the Sun, its XUV radiation and wind.

The thermospheric profiles were calculated assuming an atmosphere with the composition of the modern Earth and a zenith angle of 66 degrees. This zenith angle was assumed so that the results give approximations for the real thermospheric profiles averaged over all longitudes and latitudes, as was shown to be the case by \citeA{Johnstone18}. The profiles we adopt were calculated assuming a typical solar maximum spectrum for the modern conditions and the spectrum of an early Sun by \citeA{Claire12}.

In the Hadean and Archaean eons, the composition of the Earth atmosphere was different than today (\citeNP{Sheldon06,Feulner12,Lammer18}, and references therein) and a higher mixing ratio of CO$_2$ was present. A higher mixing ratio of CO$_2$ would lead to cooling of the upper atmosphere and a lower atmospheric escape \cite{Tian08,Lichtenegger10,Johnstone18}. Therefore, our results represent the maximum possible escape rates and can be used as a guidance if an atmosphere with a modern composition could survive as early as 3~Ga. 

The level of O$_2$ also varied a lot during the Earth's history. According to some studies, the amount of oxygen was very low prior to the first GOE at around 2.5~Ga and even after and reached a level close to the modern one not earlier than 600 Myr ago \cite{Catling05,Gebauer17}.  However, most recent studies \cite{Lyons14} indicate that oxygen rose to a nearly modern level immediately after GOE. According to \citeA{Lyons14}, then it fell again and stayed within a range of $10^{-4}$-$10^{-1}$ Present Atmospheric Level (PAL) between 2.0 and 0.5 gigayears ago. These findings show that the exact way oxygen rose on Earth has not yet been precisely identified, but it is certain that a lower oxygen mixing ratio was present prior to 0.5~Ga. Two additional simulations with 1\% and 15\% mixing ratios of oxygen 2.5 gigayears ago allow us to discuss possible implications of a lower oxygen mixing ratio for the atmospheric escape. 

\begin{figure}
\begin{center}
\includegraphics[width=1.0\columnwidth]{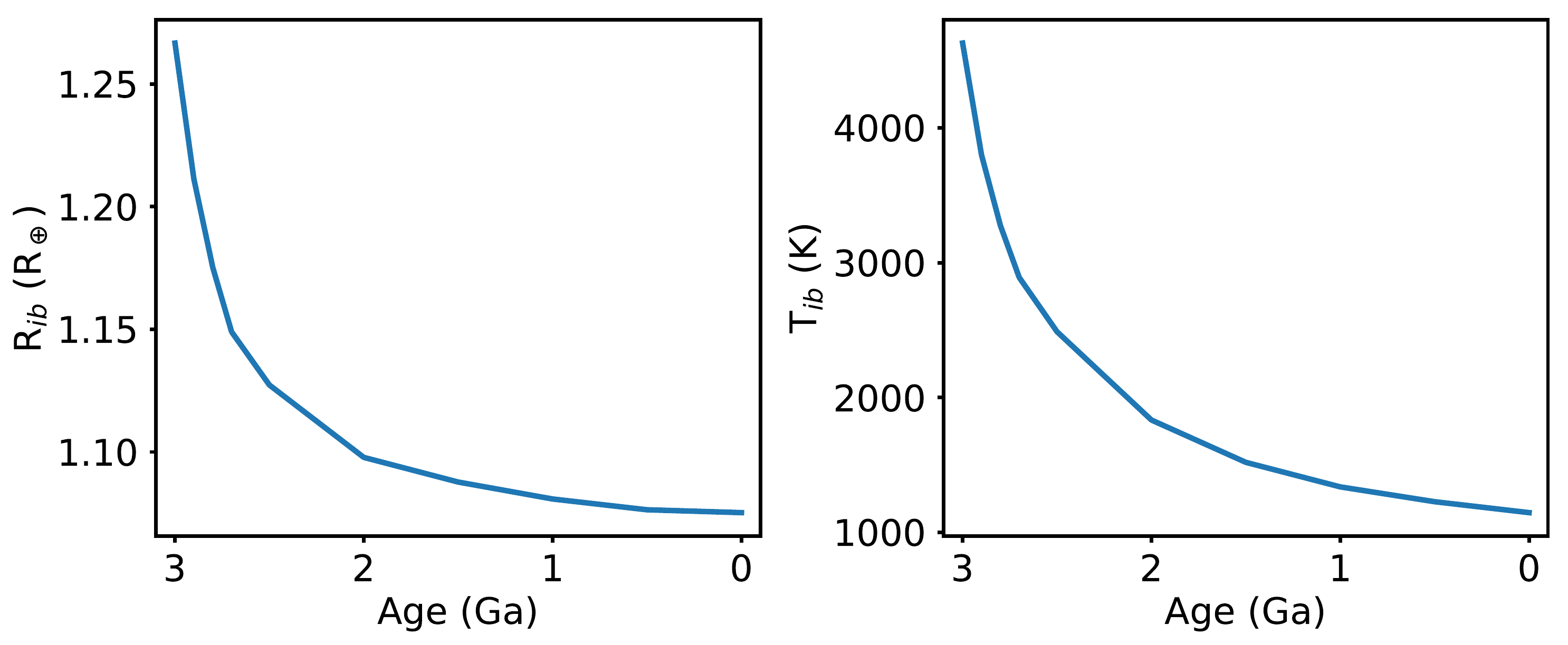}
\caption{Left panel: inner boundary location. Right panel: inner boundary temperature \cite{Johnstone18}. The present time is denoted by 0~Ga (zero gigayears ago).}  
\label{f_ib}
\end{center}
\end{figure}

We use the atmospheric profiles calculated with the Kompot code to obtain the conditions at the DSMC simulation's inner boundary. The inner boundary is chosen to lie approximately two per cent below the exobase, and the density and temperature of atomic oxygen and nitrogen calculated with the Kompot code are adopted as inputs for the DSMC model. The exobase location where the atmosphere becomes non-collisional is also found from the Kompot code profiles. The inner boundary location and temperature are shown in Fig.~\ref{f_ib}. Atmospheric parameters adopted in the simulations are summarized in Table~\ref{t_atm}. We also use the Kompot code to calculate the photoionization rates of neutral oxygen and nitrogen atoms (Table~\ref{t_atm}) in response to the changing solar spectrum.

 \begin{table}
 \caption{Atmospheric parameters adopted in the simulations. Here, R$_{ib}$ is the location of the simulation inner bondary (given in Earth's radii, R$_\oplus$), T$_{ib}$ is the inner boundary temperature (K), and n$_{atm}$ is the density of atomic oxygen and nitrogen at the inner boundary (given in $10^{12}$~m$^{-3}$). Finally, $\tau_{i}$ are the photoionization rates for atomic oxygen and nitrogen (given in $10^{-7}$~s$^{-1}$), and $\tau_{ch}$ is the charge exchange ionization rate for both atomic oxygen and nitrogen (given in $10^{-7}$~s$^{-1}$). }
 \centering
 \begin{tabular}{lccccccc}
 \hline
  Age  & $R_{ib}$  & $T_{ib}$ & $n_{atm}$ (N) & $n_{atm}$ (O) & $\tau_{i}$ (N) & $\tau_{i}$ (O)& $\tau_{ch}$ (O,N)   \\
 \hline
  0~Ga    & 1.075  & 1146 & 0.89 & 63.8 & 4.8 & 2.0 & 2.1 \\
  0.5~Ga  & 1.076  & 1227 & 1.25 & 74.3 & 5.4 & 2.2 & 2.2 \\
  1.0~Ga  & 1.080  & 1337 & 1.65 & 75.3 & 6.2 & 2.5 & 2.5 \\
  1.5~Ga  & 1.087  & 1519 & 2.55 & 78.9 & 7.3 & 3.0 & 2.8  \\
  2.0~Ga  & 1.097  & 1833 & 5.06 & 85.2 & 8.7 & 3.5 & 3.1 \\
  2.5~Ga  & 1.127  & 2488 & 9.67 & 64.3 & 11.0 & 4.4 & 3.7 \\
  2.5~Ga\footnote{1}  & 1.309 & 4519 & 11.1 & 1.07 & 11.0 & 4.4 & 3.7 \\
  2.5~Ga\footnote{2}  & 1.155 & 2864 & 9.82 & 37.1 & 11.0 & 4.4 & 3.7 \\
  2.7~Ga  & 1.148  & 2890 & 10.9 & 49.7 & 12.0 & 4.8 & 4.0 \\
  2.8~Ga  & 1.175  & 3276 & 10.4 & 35.4 & 13.0 & 5.1 & 4.1 \\
  2.9~Ga  & 1.211  & 3800 & 9.89 & 25.2 & 14.0 & 5.4 & 4.3 \\
  3.0~Ga  & 1.266  & 4629 & 10.1 & 18.4 & 15.0 & 5.8 & 4.5 \\
 \hline
 \multicolumn{2}{l}{$^{1}$One per cent mixing ratio of O$_2$}\\
 \multicolumn{2}{l}{$^{2}$Fifteen per cent mixing ratio of O$_2$}
 \label{t_atm}
 \end{tabular}
 \end{table}

\subsection{Evolution of the Earth's magnetosphere}

The Earth's magnetosphere has changed over the past few billion years due to changes in the strength of the intrinsic magnetic field and due to the evolution of the solar wind.
Measurements of the remnant magnetization in ancient rocks show that the Earth has had an active dynamo as early as 3.5 or even 4.0 billion years ago \cite{Tarduno15}. Earth's magnetic field has also experienced multiple polarity switches, during which the surface field strength decayed to about 25\% or less of its present value \cite{Glassmeier04}, however, such polarity reversals may be a feature of the last 200~Myr of the Earth's history \cite{Jellinek15}. We do not consider the effect of the recent polarity switches in this paper and focus instead on the long-time trends in the magnetic field evolution. 

Although different XUV and wind evolutionary tracks are possible for the young Sun due to different possible initial rotation rates, they converge to the same track by the time we are interested in (3~Ga and later). To determine the width of the magnetosphere and the location of the stand-off distance, we account for the evolution of the solar wind and the Earth's internal dipole. The evolution of the solar mass loss rate and wind density and velocity was studied by \citeA{Johnstone15a} and \citeA{Johnstone15b} whose results we adopt. The wind mass fluxes in their model are calculated empirically by comparing the observed rotational evolution of solar mass stars with a physical rotational evolution model. In this article, we do not account for the higher collection area of the magnetosphere in comparison to a non-magnetized planet and assume the unperturbed parameters of the solar wind in the open field line bundle of the magnetosphere. More studies on the influence of the ancient solar wind on the plasma conditions in the cusps and on the escape are necessary.
 
To calculate the magnetosphere parameters, we use the Paraboloid Magnetosphere Model (PMM). This model was developed for the Earth \cite{Alexeev06} and has been successfully applied to exoplanets \cite{Khodachenko12} and early Earth \cite{Scherf18}. The paleomagnetic field intensity for calculating the magnetosphere parameters is adopted from the study by \citeA{Biggin15}, while the solar wind input parameters were retrieved from the solar wind evolution model of \citeA{Johnstone15b}. The PMM model uses the  measurements of the paleomagnetic field strength and the ancient solar wind ram pressure as inputs and calculates the location of the magnetosphere's subsolar point $R_s$, the shape and width of the magnetosphere, the polar cap co-latitude, and the co-latitude of the equatorward boundary of the auroral oval.

The general shape of the magnetosphere is described by two parameters: these are the radius of the subsolar point, $R_s$, and the magnetosphere's width $R_t$, which is defined as the distance from the planet's center to the inner edge of the tail current sheet. The PMM calculates $R_s$ through an empirically derived formula by \citeNP{Shue97,Shue98} which is based on magnetopause crossings of different spacecraft such as ISEE 1 and 2, and Geotail: 
\begin{equation}\label{eq_shue}
  R_s = k_{p}^{1/3}(1.2\,p_{sw})^{-\frac{1}{6.6}}\mathrm{R_e} (11.4 + A B_z),
\end{equation}
where 
\begin{equation}\label{eq_shue1}
  A =   \Big\{ \begin{array}{cl}0.013& \mbox{for } B_z \geq 0 \\
  0.140 & \mbox{for } B_z < 0\end{array},
  \end{equation}
$B_z$ is the z-component of the interplanetary magnetic field, $p_{sw}$ is the solar wind ram pressure, and $k_p$ is a parameter that describes the strength of the Earth's magnetic dipole at a given time. For the simulations of the magnetosphere we assumed $B_z = 0$ in all cases.

We adopted Eq.~\ref{eq_shue} to account for the terrestrial field strength variations in the past by introducing a new parameter $k_p=M_e^{\ast}/M_e$ which is the ratio of the ancient terrestial magnetic dipole moment, $M_e^{\ast}$, to the present-day dipole moment, $M_e$. The magnetopause is described by a paraboloid of revolution \cite{Alexeev98,Alexeev06} for which the magnetospheric width $R_t$ (the magnetopause distance at the terminator) can be calculated as $R_t = \sqrt{2R_s}$. The boundary between the magnetosphere and the interplanetary space can be described by the following equation:
\begin{equation}
x = R_s \left(1 -  \frac{y^2 + z^2}{R_t^2} \right).
\label{e_mf}
\end{equation}
Here, the coordinates are defined as follows: the $x$-axis points towards the Sun, the $y$-axis points into the opposite direction to the Earth's motion, and the $z$-axis completes the right-hand coordinate system. In addition to the paraboloid magnetosphere, we also introduce the open field line bundle, which are conic-shaped regions near the planet's poles within the polar opening angle. The latter depends on the magnetosphere's stand-off distance, which in its turn is determined by the strength of the internal dynamo and the external dynamic pressure of the solar wind and can be estimated as \cite{Alexeev06}

\begin{equation}
\theta = \sqrt{0.52 R_\oplus/R_s}.
\label{e_angle}
\end{equation}

\begin{figure}[t]
\begin{center}
\includegraphics[width=1.0\columnwidth]{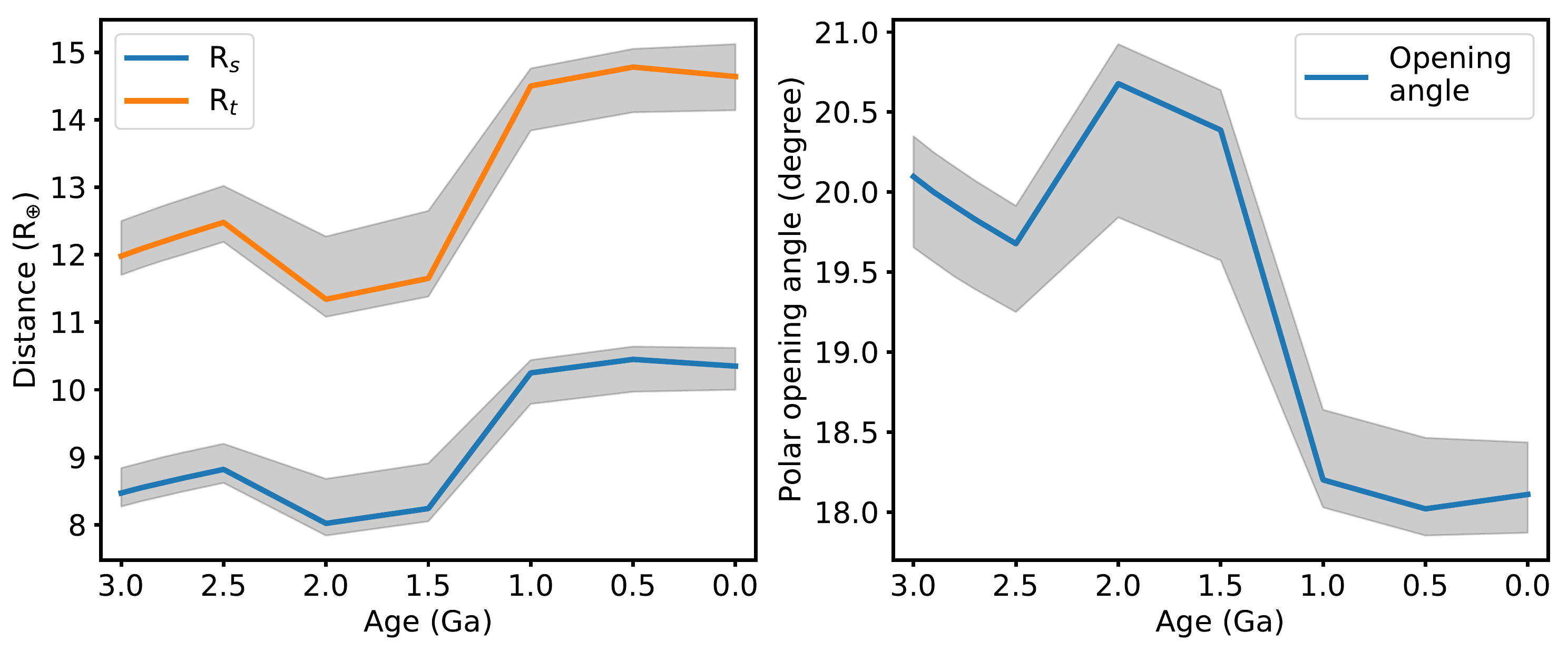}
\caption{Left panel: Evolution of the location of the subsolar point (R$_s$) and width (R$_t$) of the Earth's magnetosphere in time. The solid blue and orange lines show the average values. The shaded area illustrates the maximum and minimum values. Right panel: average (solid line) and minimum and maximum (shaded area) polar opening angle.}  
\label{f_magnetosphere}
\end{center}
\end{figure}

This angle indicates the location of the last closed magnetic field line, i.e., the last line that originates and closes at the Earth. The coefficient 0.52 accounts for a part of the open field flux that eventually closes at the night side of the magnetosphere in the equatorial region, thus reducing the open field flux, as shown in numerical integration of the flux in the PMM \cite{Belenkaya08,Belenkaya17}. The angle $\theta$ is shown in Fig.~\ref{f_clouds}. This formula does not account for the asymmetries of the magnetosphere and for this reason may slightly overestimate the size of the magnetospheric region with the open field lines \cite{Egan19}. Eq.~\ref{e_angle} describes the opening angle correctly for strongly magnetized planets, when the magnetosphere's stand-off distance lies well above the ionosphere. Indeed, the Earth was always in the strongly magnetized regime during the time period that we consider \cite{Egan19}. 
 
Fig.~\ref{f_magnetosphere} illustrates the evolution of the magnetosphere's stand-off distance, width and the polar opening angle that we adopt in our simulations. The maximum values of $R_s$ and $R_t$ (and minimum values for the opening angle) are for the slow rotator wind and the maximum paleomagnetic field intensity (according to \citeNP{Biggin15}), whereas the minimum is for the fast rotator wind and the minimum paleomagnetic field intensity. The solar wind parameters for fast and slow rotator are calculated according to \citeA{Johnstone15b}. One can see that although the opening angle increases as one goes back in time, the difference to the modern magnetosphere is not extreme. In fact, the short-term variability due to CME impacts that we do not take into account for can exceed this difference \cite{Airapetian16}. It is possible that the CME impacts were more frequent many gigayears ago that led to temporary larger polar opening angles than those estimated by Eq.~\ref{e_angle} \cite{Airapetian16,Kay19}, but in this article we focus on the long-term trends and do not take into account the influence of the CMEs. The magnetosphere's parameters adopted in the simulations are summarized in Table~\ref{t_wind}.

The same solar wind parameters that we use \cite{Johnstone15b} were adopted in a recent study by \citeA{Carolan19} who have also investigated the shape and size of the ancient Earth's magnetosphere. They used the MHD model BATS-R-US which is part of the Space Weather Modeling Framework \cite{Toth05} and arrived at very similar conclusion to ours regarding the magnetosphere's width and stand-off distance. However, they did not account for the evolution of the internal dipole of the Earth. The general agreement of the two different approaches proves the robustness of our results.

 \begin{table}
 \caption{Parameters of the solar wind and magnetosphere adopted in the simulations. Here, $k_p = M_e^{\ast}/M_e$ is the   paleomagnetic field intensity as compared to today's Earth (adopted from \citeNP{Biggin15}), R$_s$ and R$_t$ are the distance to the magnetosphere subsolar point (given in R$_\oplus$) and magnetosphere width (given in R$_\oplus$), respectively. Furthermore, $n_{sw}$, $v_{sw}$, and $T_{sw}$ are the solar wind density (given in $10^6$~m$^{-3}$), velocity (km~s$^{-1}$), and temperature ($10^4$~K), respectively. The polar opening angle $\theta$ is given in degrees.}
 \centering
 \begin{tabular}{lccccccc}
 \hline
  Age  & $M_e^{\ast}/M_e$  & $R_s$ & $R_t$ & $n_{sw}$ & $v_{sw}$ & $T_{sw}$ & $\theta$  \\
 \hline
  0~Ga    & 1.0  & 10.35 & 14.64 & 4.67 & 448  & 8.90  & 18.10  \\
  0.5~Ga  & 1.08 &   10.45 & 14.78 & 4.90 & 461  & 9.30  & 18.01 \\
  1.0~Ga  & 1.08 & 10.25 & 14.50 & 5.20 & 478  & 9.77  & 18.20   \\
  1.5~Ga  & 0.6  &  8.24 & 11.65 & 5.57 & 497  & 10.35 & 20.38 \\
  2.0~Ga  & 0.6  &  8.02 & 11.34 & 6.04 & 521  & 11.09 & 20.67 \\
  2.5~Ga  & 0.88 &  8.82 & 12.48 & 6.66 & 553  & 12.05 & 19.67 \\
  2.7~Ga  & 0.88 &  8.69 & 12.29 & 6.96 & 568  & 12.53 & 19.82 \\
  2.8~Ga  & 0.88 &  8.62 & 12.19 & 7.13 & 576  & 12.79 & 19.91 \\
  2.9~Ga  & 0.88 &  8.55 & 12.09 & 7.31 & 586  & 13.08 & 19.99 \\
  3.0~Ga  & 0.88 &  8.47 & 11.98 & 7.50 & 595  & 13.37 & 20.09 \\
 \hline
 \label{t_wind}
 \end{tabular}
 \end{table}

\subsection{Exosphere model and the polar outflow escape rate}

For the exosphere, we use a Direct Simulation Monte Carlo (DSMC) kinetic model designed to simulate the interaction between the planet's atmosphere and the stellar or solar wind. The code follows particles in a 3D simulation domain and includes multiple neutral and ion species. It considers the main processes acting on an atom in a planetary exosphere, such as ionization by the central star's radiation and wind, charge exchange between the wind protons and the atmospheric neutrals, radiation pressure, gravity, and elastic collisions between the neutral particles. The model has been applied to various types of planetary atmospheres, including hydrogen-dominated atmospheres \cite{K14a,K14b}, early steam atmospheres \cite{Lichtenegger16}, and secondary atmospheres such as those dominated by nitrogen \cite{K19}. The detailed description of the code is provided in \citeA{Holmstroem08} and \citeA{K14a}.

The version of the code that we apply here includes neutral and ionized nitrogen and oxygen particles. Each of the particles (``metaparticles'') simulated in the code represents a large amount of the real particles that are all considered to be moving together. The metaparticles representing neutral oxygen and nitrogen atoms are launched at a random location at the spherical inner boundary that is shown in Fig.~\ref{f_ib}. The random initial velocity of each metaparticle is drawn from a Maxwellian distribution according to the atmosphere's temperature as described in \citeA{Holmstroem08,K14a}. Then, the metaparticles are followed in the simulation domain that had the size of $8 \times 10^8$~m in all directions. 

As stated above, in addition to a paraboloid magnetosphere that was adopted as the magnetosphere boundary in earlier articles (e.g., \citeNP{K19}), we account for the presence of the open field line bundle with an opening angle described by Eq.~\ref{e_angle}. In our set of simulations, neutral oxygen and nitrogen atoms can undergo photo- and charge exchange ionization only if they are located within the polar opening angle described by Eq.~\ref{e_angle} near both poles of the Earth. The amount of ionizations per second is then calculated and used as a proxy for the polar outflow escape rate, which is a reasonable assumption since it is the available ionospheric ion density that limits the outflow mechanism on the Earth \cite{Andre15}. We assume that only particles that underwent ionization can escape along the open field lines, because they can be accelerated by the local electric fields and follow the open field lines away from the planet. Depending on their energy, these ions can end up in different regions of the Earth's magnetosphere, such as the magnetopause, the distant tail, and the ring currents, or fall down to the atmosphere \cite{Ebihara06}. Since we are primarily interested in the upper limits of the escape to understand if the nitrogen-dominated atmosphere of the Earth can survive the atmospheric escape in the considered time span, such a simplification seems justified. In future studies, we will account for the Lorentz force and the gyration of the ions to further constrain the upper limit of escape, which is beyond the scope of the present article.

\begin{figure}
\begin{center}
\includegraphics[width=1.0\columnwidth]{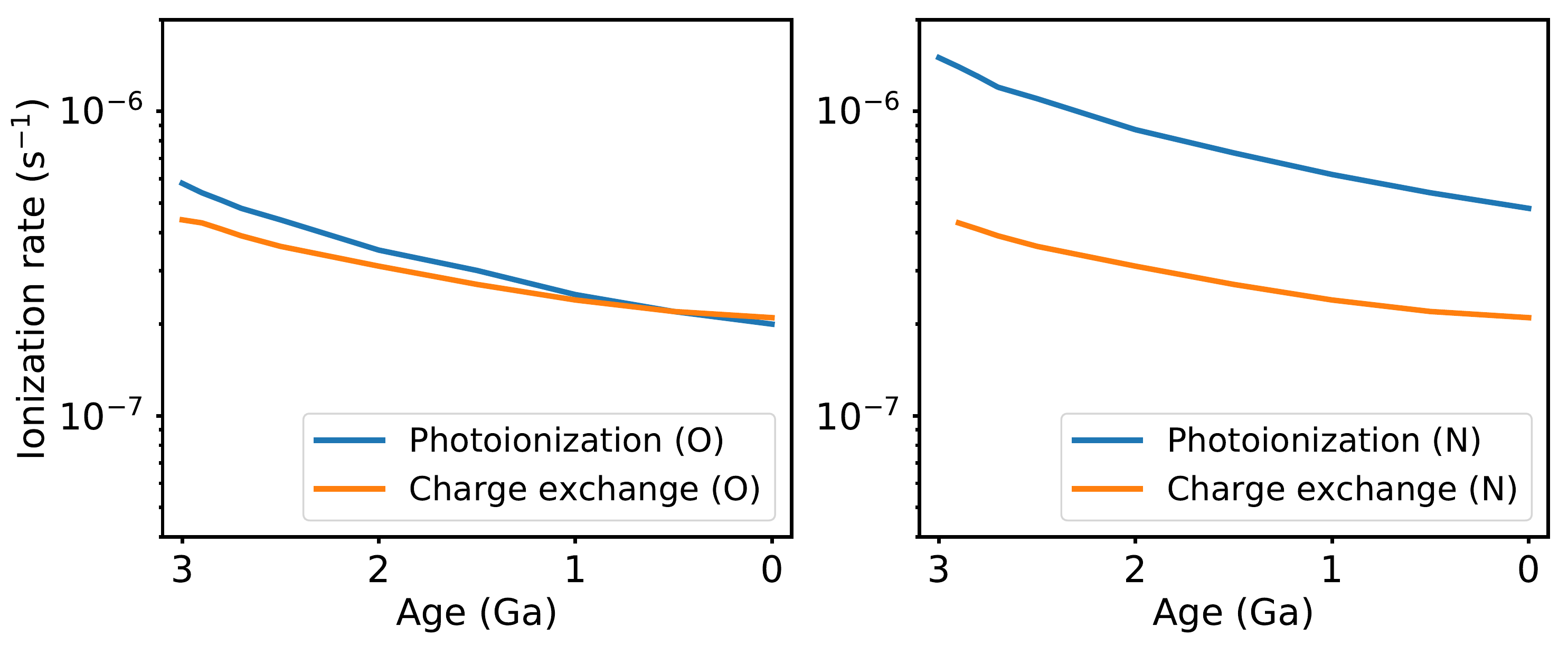}
\caption{Ionization rates of neutral oxygen (left panel) and nitrogen (right panel) atoms due to charge exchange and photoionization.}  
\label{f_ionrates}
\end{center}
\end{figure}

Solar protons are not included in the simulation as particles, but are replaced by the charge exchange ionization rate defined as
\begin{equation}
\tau_{ch} = n_{sw}\sigma_{ch}v_{sw},
\label{e_chrate}
\end{equation}
where $n_{sw}$ is the solar wind density, $\sigma_{ch}$ is the corresponding charge exchange cross section, and $v_{sw}$ is the speed of the solar wind. Since the atmosphere is hydrostatic in all cases we consider, the bulk outflow velocity of the atmosphere equals zero, and we only account for the speed of the solar wind in Eq.~\ref{e_chrate}. Figure~\ref{f_ionrates} illustrates the evolution of photoionizaion and charge exchange ionization rates for atomic oxygen and nitrogen in time. Charge exchange rates for the two species are equal because we assume the same collision cross section for both of them based on the measurements \cite{Lindsay05,Rinaldi11}. As a simplification, we assume the non-shocked parameters of the solar wind to calculate the charge exchange ionization rate in Eq.~\ref{e_chrate}. The particles can also be photoionized. The electron impact ionization plays no significant role for the conditions considered here due to insufficient density and temperature of the solar wind. We adopt the following cross section for the charge exchange reaction between nitrogen and oxygen neutrals and solar wind protons: $\sigma_{ch} = 10^{-19}$~m$^2$ \cite{Lindsay05,Rinaldi11}.

Following \citeNP{K19}, we estimate elastic collision cross sections between neutral species \cite{Atkins00} as
\begin{equation}
	\sigma_{a-b} = \pi (R_a + R_b)^2,
\end{equation}
where $\sigma_{a-b}$ is the elastic collision cross section between the two species $a$ and $b$, and $R_a$ and $R_b$ are the atomic radii of corresponding species. We adopt the following elastic collision cross sections in our simulations: $\sigma_{O - O} = 2.9 \times 10^{-20}$~m$^{2}$, $\sigma_{N - N}  = 3.9 \times 10^{-20}$~m$^2$, $\sigma_{N - O} = 3.4 \times 10^{-20}$~m$^2$. 	

Since we are interested mainly in the production rates of oxygen and nitrogen ions, the gravity force is applied only to the neutral atoms and not to the newly produced ions that are allowed to move freely through the simulation domain. Oxygen and nitrogen neutral atoms are ionized only if they are located in the polar open field line bundle. We define an open field line bundle as the area located within the magnetosphere described by Eq.~\ref{e_mf} and within the polar angles described by Eq.~\ref{e_angle}. In summary, we have two sources of ionization in the open field line regions: photoionization and charge exchange. Observations show that solar wind particles do reach the Earth's cusps and precipitate down the magnetic field lines \cite{Chen05,Parks08}. It is likely that not all solar wind protons reach the exobase, but since we are interested in the upper layers of the atmosphere (exosphere), we assume the medium to be optically thin, so that the photoionization rate and the charge ionization rate can be kept constant. Previous works have shown that charge exchange with the solar wind happens high in the upper layers of the atmosphere and exosphere and can accelerate the escape \cite{Chassefiere96,K13,Lichtenegger16}. The ionization rates are summarized in Table~\ref{t_atm}. Not all newly produced ions will escape, therefore, our results should be treated as an upper limit for the outflow rates.

\section{Results}
\label{sec_results}

\begin{figure}
\begin{center}
\includegraphics[width=1.0\columnwidth]{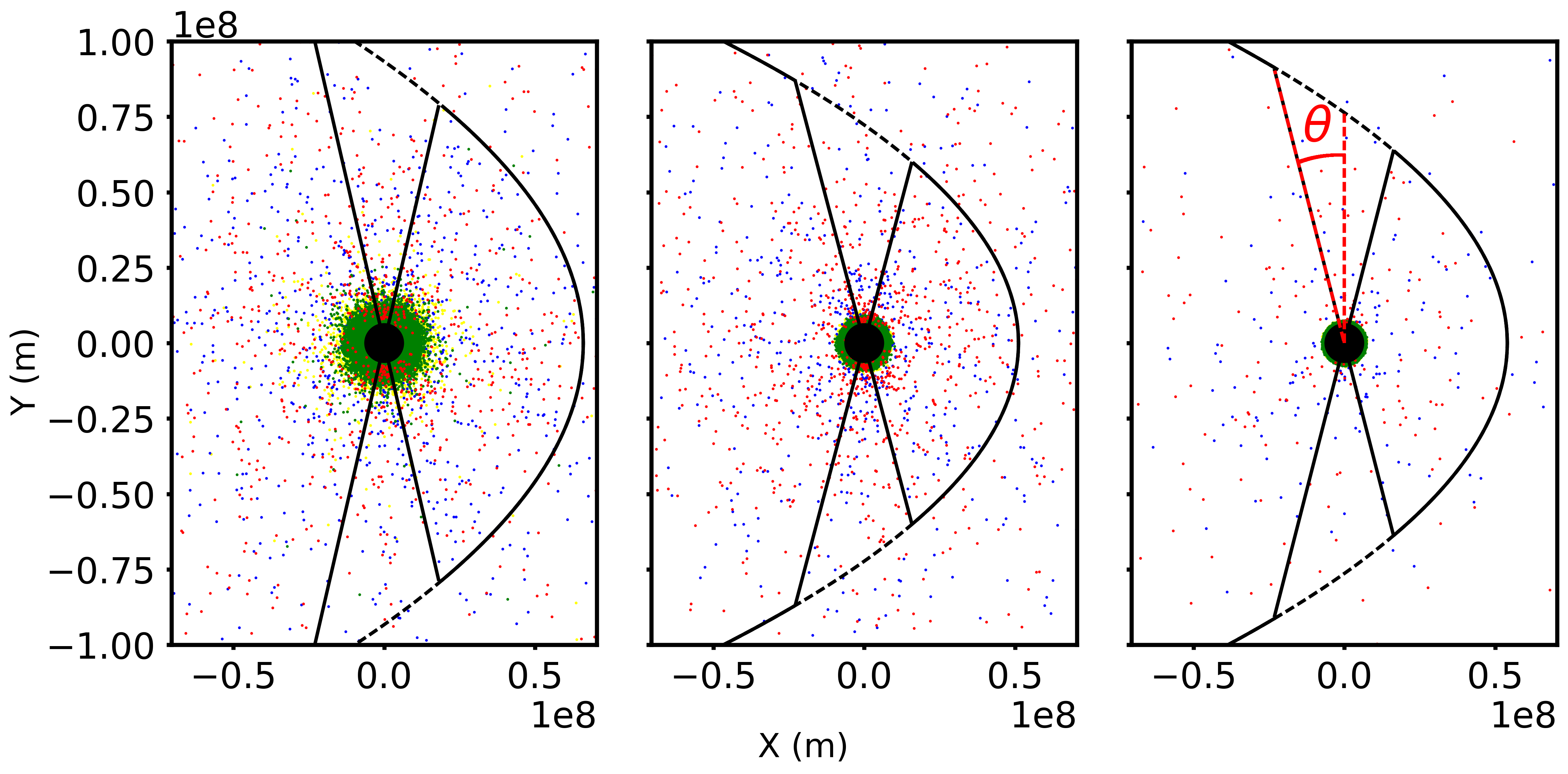}
\caption{Modeled atomic clouds surrounding the Earth at different times (from left to right: 3.0~Ga, 2.5~Ga, present-day). The solid black line shows the shape of the magnetosphere, the dashed lines show the locations of the open field line bundle. The colored dots are the simulated metaparticles representing atmospheric atoms and ions. Neutral nitrogen atoms are yellow, neutral oxygen atoms are green, nitrogen ions are blue, and oxygen ions are red. The polar opening angle $\theta$ is shown in the right panel.}        
\label{f_clouds}
\end{center}
\end{figure}

Fig.~\ref{f_clouds} shows three example simulations of the exosphere and open field line bundle of the Earth at its present age, and at 2.5 and 3~Ga. Going back in time, the polar opening angle increases, reaching a maximum at approximately 2~Ga (Fig.~\ref{f_magnetosphere}), and the magnetosphere sub-stellar point moves closer to the planet's surface due to the higher dynamic pressure of the solar wind and weaker magnetic field. However, as one can see from Fig.~\ref{f_clouds}, the differences in the shape of the magnetosphere is not significant enough for the subsolar point to reach the upper atmosphere's boundary even three gigayears ago. The structure of the upper atmosphere undergoes significant changes. At present, the exobase is located very close to the Earth's surface at $\sim$500~km (see right panel of Fig.~\ref{f_clouds}). Going back in time, the exobase location moves upwards due to the expansion caused by a higher level of the solar XUV radiation. One can see it in the increasing number of green (oxygen) and yellow (nitrogen) neutral particles further from the Earth, especially 3~Ga. The subsolar point of the magnetosphere never comes close enough to the Earth's surface to allow for direct interactions between the solar wind and the Earth's ionosphere, meaning that escape takes place in the open field line bundle during the entire studied time period. This is not the case for lighter hydrogen atoms that can reach the subsolar point, but in this study we focus on the heavier oxygen and nitrogen atoms. The number of ionized particles also increases from the present-day to 3~Ga. We do not take into account the escape of O$_2^+$ and N$_2^+$ and only focus on atomic oxygen and nitrogen that are the most common ions in the polar regions and provide the  major contribution to polar outflow \cite{Yau93}.

\begin{figure}
\begin{center}
\includegraphics[width=0.5\columnwidth]{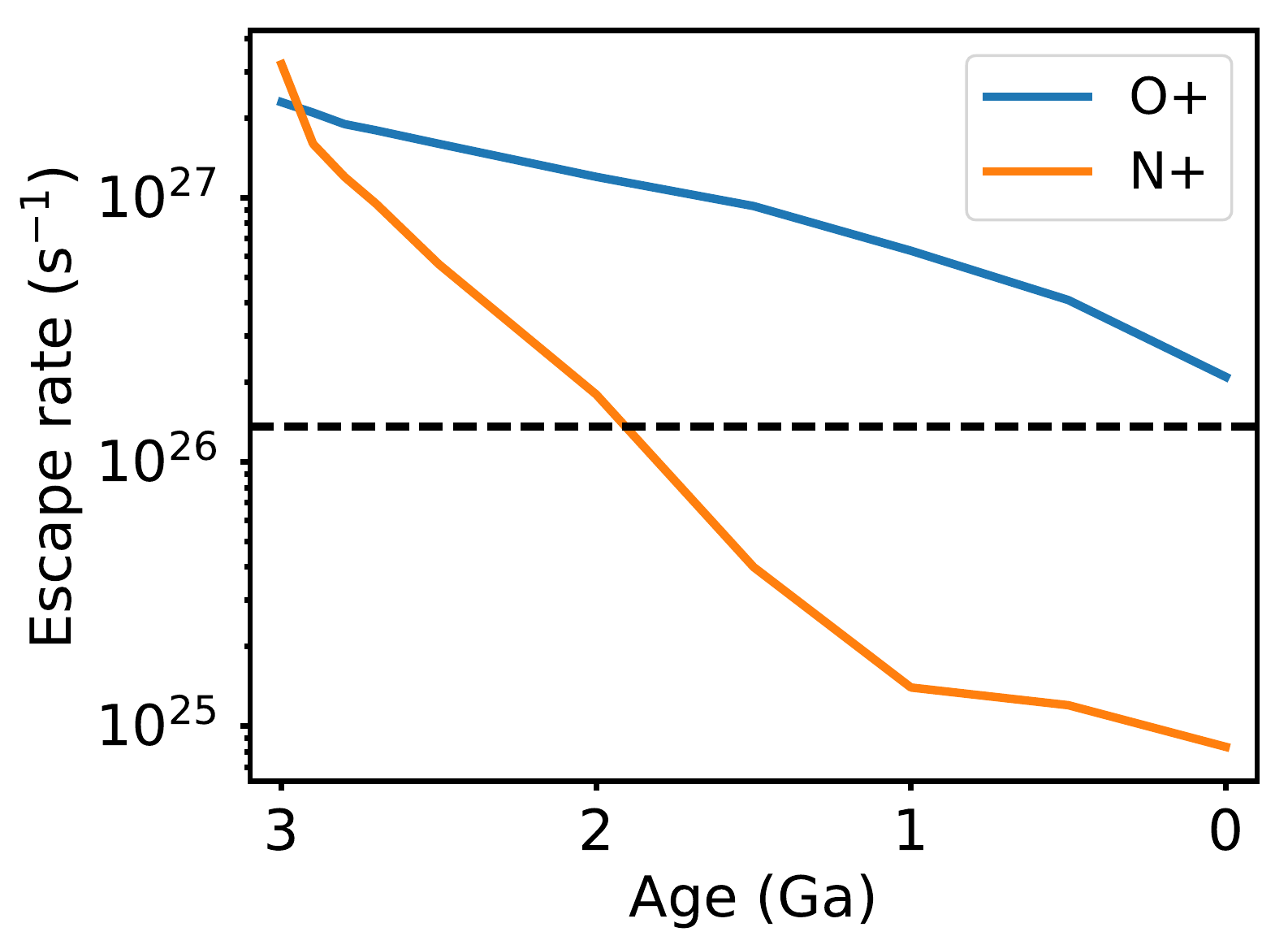}
\caption{Evolution of the maximum possible polar outflow loss rates from the Earth, assuming the present-day atmospheric composition. The dashed line shows the present-day measured escape rate of oxygen ions during high solar activity of $1.36 \times 10^{26}$~s$^{-1}$ \cite{Schillings19}. As one can see, going back in time the escape rate of nitrogen ions is increasing faster than oxygen ions. } 
\label{f_esc}
\end{center}
\end{figure}

Fig.~\ref{f_esc} illustrates the evolution of the maximum polar outflow escape rates for O$^+$ and N$^+$ ions. Our model predicts an escape rate of $2.1\times10^{26}$ oxygen ions per second for modern conditions. This number is the production rate of oxygen ions in the open field line bundle due to charge exchange and photoionization. Not every produced ion escapes the polar region; depending on their initial velocity and location, these ions might end up in different regions of the Earth's magnetosphere or even fall down into the atmosphere, if their energy is low enough. \citeA{Ebihara06} showed that the majority (more than 90\% depending on their initial energy) of newly produced oxygen ions are lost from the atmosphere and end up in the different regions of the magnetosphere. Their further fate can differ depending on their energy and it is not completely clear how many of them actually escape.  Some studies indicate that most of these ions can actually be recaptured by the Earth \cite{Seki01,Haaland12}, but other authors (e.g., \citeNP{Nilsson12,Schillings19}) claim that the majority of these ions are permanently lost. Some results \cite{Engwall09} show that even initially cold ions travel along the magnetic field lines and significantly contribute to the Earth's polar outflow.

According to observations, the loss rate of oxygen ions due to polar outflow varies in the range of $1 \times 10^{24}-1.6\times 10^{26}$~s$^{-1}$, depending on the wind conditions \cite{Slapak17,Schillings19}. The escape rate for the modern Earth that we estimate is $2.1 \times 10^{26}$~s$^{-1}$ which is in a good agreement with a value during high solar activity and solar wind dynamic pressure \cite{Slapak17,Schillings19}. One can expect that because the atmospheric profiles by \citeA{Johnstone18} that we adopted were calculated for a high solar activity case. The observed flux of nitrogen ions is comparable to the flux of ionized oxygen, although often it is slightly (up to a factor of two) lower \cite{Yau93}, which is also in agreement with our estimates. We conclude that our results correctly reproduce the observed polar outflow rates at present and can be used as a proxy for the past escape rates. However, they should be treated as an upper limit of the real escape, because some of these particles may be later recaptured by the atmosphere. The calculated escape rates are summarized in Table~\ref{t_escape}.

 \begin{table}
 \caption{Estimated maximum polar outflow escape rates for oxygen and nitrogen ions from both open field line areas in particles per second ($s^{-1}$). Integrated escape rates from 3~Ga to present: $2.0 \times 10^{18}$~kg or 39\% atmosphere mass for oxygen and $5.25 \times 10^{17}$~kg or 10\% atmosphere mass for nitrogen, assuming the total atmosphere mass of $5.148 \times10^{18}$~kg. }
 \centering
 \begin{tabular}{l c c}
 \hline
  Age  & O$^+$ &  N$^+$ \\
 \hline
  0.0~Ga  & $2.1 \times 10^{26}$ & $8.4 \times 10^{24}$ \\
  0.5~Ga  & $4.2 \times 10^{26}$ & $1.2 \times 10^{25}$ \\
  1.0~Ga  & $6.3 \times 10^{26}$ & $1.4 \times 10^{25}$ \\
  1.5~Ga  & $9.3 \times 10^{26}$ & $4.0 \times 10^{25}$ \\
  2.0~Ga  & $1.2 \times 10^{27}$ & $1.8 \times 10^{26}$ \\
  2.5~Ga  & $1.6 \times 10^{27}$ & $5.6 \times 10^{26}$ \\
  2.5~Ga\footnote{3}  & $1.0 \times 10^{26}$ & $2.9 \times 10^{27}$ \\  
  2.5~Ga\footnote{4}  & $1.1 \times 10^{27}$ & $7.5 \times 10^{26}$ \\
  2.7~Ga  & $1.8 \times 10^{27}$ & $9.5 \times 10^{26}$ \\
  2.8~Ga  & $2.0 \times 10^{27}$ & $1.3 \times 10^{27}$ \\
  2.9~Ga  & $2.2 \times 10^{27}$ & $1.6 \times 10^{27}$ \\
  3.0~Ga  & $3.3 \times 10^{27}$ & $2.3 \times 10^{27}$ \\
 \hline
 \multicolumn{2}{l}{$^{3}$One per cent mixing ratio of O$_2$}\\
 \multicolumn{2}{l}{$^{4}$Fifteen per cent mixing ratio of O$_2$ }
 \label{t_escape}
 \end{tabular}
 \end{table}

The total integrated losses for oxygen and nitrogen amount to $2.0 \times 10^{18}$~kg and $5.3 \times 10^{17}$~kg, or 39\% and 10\% of the modern atmosphere's mass, respectively, where we assume the atmosphere's mass equal to $5.148 \times 10^{18}$~kg. For oxygen, the escape is especially significant, being high enough to remove large fractions of the modern atmospheric oxygen reservoir and might have potentially influenced the evolution of oxygen in the Earth's atmosphere. The strong escape rate of oxygen is mostly a result of its high density in the upper atmosphere. At present, atomic oxygen is the dominant species near the Earth's exobase. Although the detailed study of the evolution of oxygen escape is outside the scope of the present study, our simulation with 1\% mixing ratio at 2.5~Ga predicts a much lower oxygen escape rate of only $1.0 \times 10^{26}$~s$^{-1}$. The simulation with 15\% mixing ratio of oxygen predicts a loss rate of oxygen ions of $1.1 \times 10^{27}$~s$^{-1}$, which is expectedly slightly lower than the oxygen escape rate obtained assuming the modern composition 2.5~Ga. This result indicates that the atmospheric escape depends significantly on the atmosphere's composition.

\begin{figure}
\begin{center}
\includegraphics[width=0.8\columnwidth]{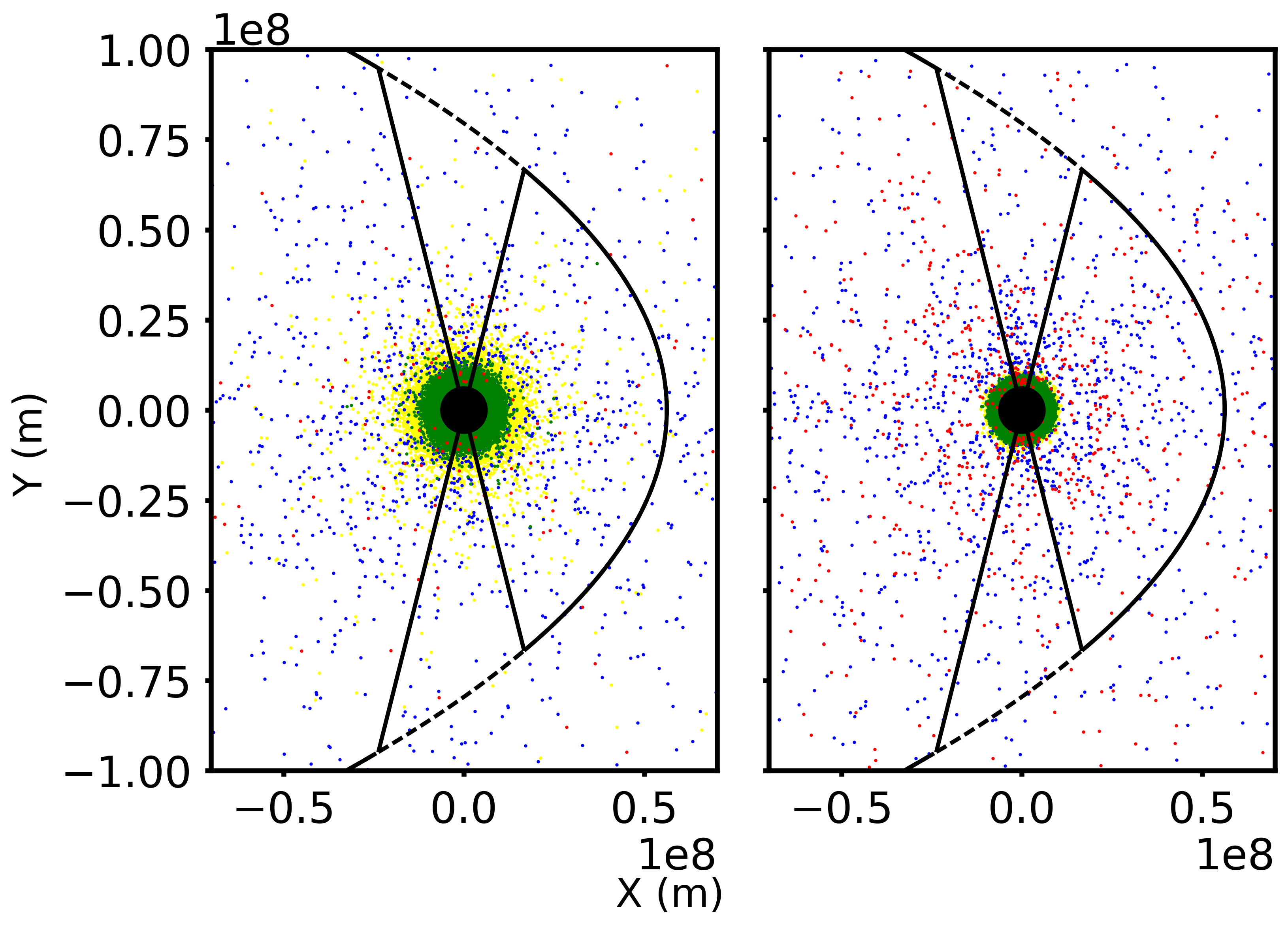}
\caption{Modeled atomic cloud around the Earth 2.5~Ga assuming a one per cent (left panel) and a fifteen per cent (right panel) mixing ratios of oxygen in the atmosphere. The colors are the same as in the Fig.~\ref{f_clouds}. The amount of oxygen ions is decreased in comparison to the middle panel of Fig.~\ref{f_clouds}, and the atmosphere is more expanded compared to the 2.5~Ga case with modern composition, especially for the case with one per cent mixing ratio.}        
\label{f_cloud_GOE}
\end{center}
\end{figure}

At the exobase of the present-day Earth, atomic nitrogen has a much lower density than atomic oxygen and is present mostly in its molecular form \cite{Johnstone18}. Escape of molecular nitrogen is much less efficient than the escape of ionized nitrogen atoms \cite{Yau93}. Going back in time, the composition at the exobase changes between 1.5 and 2.0~Ga, when atomic nitrogen becomes more abundant than molecular nitrogen. The increase in the exobase density, combined with a higher photoionization rate for nitrogen, leads to the escape rate of nitrogen growing faster as one goes back in time in comparison to oxygen. In the mid-Archean around 3~Ga, the nitrogen escape surpasses the oxygen escape and reaches a high number of $2.3 \times 10^{27}$~s$^{-1}$. Such an escape rate would remove the entire reservoir of atmospheric nitrogen in approximately 10~Gyr, which shows that it is sustainable over a geologically long time. Since we likely overestimate the escape rate, this indicates that a nitrogen-dominated atmosphere was possible starting from 3~Ga onwards. The ion escape rates that we obtain at three billion years ago exceed the thermal Jeans escape rates by several orders of magnitude (see \citeNP{Johnstone18}, Fig. 13), which shows that non-thermal escape processes were more efficient at removing the Earth's atmosphere from this time onward. Our escape rates for oxygen are in general agreement with those estimated by \citeNP{Slapak17} for several billion years ago. Prior to this time, a higher mixing ratio of CO$_2$ or another greenhouse gas serving as an upper atmosphere coolant was likely necessary to protect the atmosphere from rapid thermal escape. This result is in agreement with climate studies \cite{Feulner12}, studies of the thermal escape and on isotope fractionation \cite{Lammer20}, and with geological evidences \cite{Catling20}.

One should also note that the escape rate of nitrogen is significantly higher at 2.5~Ga if we assume a 1\% mixing ratio of oxygen in comparison to the case with the modern atmospheric composition. In this case, it reaches $2.9 \times 10^{27}$~s$^{-1}$ which is a factor of five higher than the nitrogen escape rate at 2.5~Ga with modern composition of the atmosphere. Such escape rate would remove an amount of nitrogen equal to the mass of the Earth's atmosphere in 2.4~Gyr. Escape rate of nitrogen ions is also higher in the simulation with 15\% mixing ratio of oxygen. In this case, it increases moderately by a factor of 1.3 and reaches $7.5 \times 10^{26}$~s$^{-1}$. In both cases, the escape rate increases due to a higher atmospheric temperature and a larger expansion of the atmosphere. This is likely due to the importance of atomic oxygen, produced by the photodissociation of O$_2$, for the cooling of the upper mesosphere and thermosphere \cite{Roble87}. Throughout the middle and upper thermosphere, atomic O emits significantly at 63 and 147~$\mu$m, and in the lower thermosphere and upper mesosphere, it is the species mostly responsible for collisionally exciting CO$_2$ and NO, driving infrared cooling from these molecules. One can see that the atmosphere 2.5~Ga is significantly hotter and more expanded if the mixing ratio of oxygen is decreased (see Table~\ref{t_atm}) which results in a higher escape rate for nitrogen. As expected, the escape rate of oxygen is much lower if one assumes a 1 \% mixing ratio due to a lower density of atomic oxygen at the inner boundary, and lies slightly below the modern level. Escape rate for a 15\% mixing ratio of oxygen is slightly lower than the one calculated for the modern atmospheric composition 2.5~Ga. From these additional simulations, we can conclude that the escape rates of nitrogen can be even higher in absence of oxygen due to the decrease in cooling of the upper atmosphere, which supports the likelihood of a higher level of CO$_2$ in the Archean ($\sim$3~Ga). 

\section{Conclusions}
\label{sec_conslusions}

We have estimated the upper limits for the polar outflow escape rates of oxygen and nitrogen ions assuming the modern atmospheric composition from 3~Ga to present. In addition, we have performed two additional simulations at 2.5~Ga with mixing ratios of oxygen of 1\% and 15\% to account for different possible conditions during the rise of oxygen on Earth (GOE). We have shown that the escape rates are sensitive to the level of the XUV radiation of the early Sun and to the atmosphere's composition. 

We conclude that a modern nitrogen-dominated atmosphere could have survived the exposure to the XUV radiation of the young Sun from 3.0~Ga onwards. The total maximum loss of the nitrogen reservoir of $5.3 \times 10^{17}$~kg, or 10\% of the modern atmosphere's mass, would possibly leave a minor imprint in the isotope fractionation (e.g., \citeNP{Furi15}). If the actual escape rate of nitrogen was lower than our estimate, the atmosphere of the Earth was not significantly fractionated from three Ga onwards and the present fractionation level was reached prior to that time.

The total estimated escape rate of oxygen of $2.0 \times 10^{18}$~kg, or 39\% of the modern atmosphere's mass, is high and would deplete the atmospheric reservoir. However, since this rate was estimated assuming the present-day atmospheric composition, it likely overestimates the escape rate of oxygen. As expected, in the simulations with 1\% and 15\% mixing ratios of oxygen the estimated escape rates of oxygen ions were lower. If the modern-day mixing ratio of oxygen was not reached until much later in the Earth's history \cite{Catling05}, then the late GOE has possibly protected the oxygen reservoir from depletion. If, on the other hand, the level of oxygen was high immediately after GOE \cite{Lyons14}, then the efficient removal of oxygen by the polar outflow could have played a role in the following drop of the oxygen mixing ratio in the time period between 2.0 and 0.5 gigayears ago. 

The majority of the atmospheric ions produced in the polar regions end up in the magnetosphere \cite{Ebihara06}, but it is not clear how many are eventually recaptured by the Earth (see discussion in \citeNP{Seki01,Engwall09,Haaland12,Nilsson12,Schillings19}). In general, ions with higher energies have higher probability to escape, but the escape probability also depends on the solar activity and effectiveness of energization \cite{Slapak18,Krcelic20}. Future observations and modeling will help to better understand the global picture of the fate of ions of different species and with different initial energies. However, even taking into account that some of the escaping ions may be recaptured, the polar ion outflow escape rate exceeds the thermal escape rate during the considered time period by several orders of magnitude, which shows that the Earth's atmospheric escape was mostly influenced by non-thermal escape processes from 3~Ga to present. 
In a future study, we plan to revisit the problem and account both for the magnetospheric dynamics that guides the motion of ions, as well as the change in atmospheric composition such as the gradual increase of the oxygen mixing ratio in the atmosphere. 

An interesting feature of simulations at 2.5~Ga with 1\% and 15\% mixing ratios of oxygen is a hotter temperature, which drives the exobase higher and leads to the expansion of the atmosphere. This leads to an enhanced escape rate of nitrogen in comparison to the case 2.5~Ga with the modern composition. Therefore, the evolution of the Earth's nitrogen and oxygen atmospheric reservoirs is tangled. If atmospheric nitrogen escaped at a rate of $4 \times 10^{27}$~s$^{-1}$ as estimated assuming 1\% mixing ratio of oxygen, the atmosphere would be lost within approximately one Gyr, unless the majority of newly produced ions would be recaptured by the magnetosphere. This is an indication that the amount of CO$_2$ in the atmosphere was higher at 2.5-3.0~Ga in comparison to the current composition. This is in agreement with climate studies on the Faint Young Sun Paradox and with geological evidence (e.g., \citeNP{Feulner12,Catling20}). Future studies on nitrogen isotope fractionation can shed light on the history of the nitrogen escape and the rise of oxygen. One should also note that a higher level of CO$_2$ can lead to an increased amount of oxygen in the ionosphere produced by photodissociation, as is the case for modern Mars and Venus which have oxygen-dominated ionospheres \cite{Lundin11}. Therefore, the escape rates of atmosphere's main species are sensitive to the species' mixing ratios and should be investigated for various possible atmosphere's compositions.

Finally, we conclude that starting from 3.0~Ga onwards the polar outflow escape was governed by the evolution of the XUV output of the Sun and the atmosphere's composition and to a lesser extent by the evolution of the Earth's magnetic field, under the assumptions about the plasma parameters in the polar regions made in our model. The influence of the changing plasma environment on the escape should be investigated separately. This is due to the fact that the magnetospheric stand-off distance always stayed well above the ionosphere, so that escape always took place through the open field line bundle. This might be different prior to 3~Ga when the atmosphere was possibly even more expanded and the stand-off distance closer to the planetary surface, but in the considered time period the evolution of the magnetosphere played a lesser role for the ion outflow than the evolution of the atmosphere and the Sun's XUV output. 

Our results may be of interest also for studies on the early Mars and Venus. The remnant magnetization indicates that Mars has had an active dynamo for the first $\sim$500 million years of its history \cite{Acuna98,Lillis13}. It is interesting to wonder if the cessation of the dynamo has accelerated the loss of volatiles from the planet, making Mars the arid world it is today (e.g., \citeNP{Alho15,Kite19}). Recent results indicate that an intrinsic magnetic field has been possible also on ancient Venus \cite{ORourke18}. Current escape rate of oxygen from Venus, if extrapolated into the past, could have removed 0.02-0.6~m of a global equivalent of water \cite{Persson20}. However, if Venus indeed has had a magnetic field in the past, then a dominant escape mechanism was likely different than today, which would change the picture of its atmospheric evolution. Future studies on the comparison of escape rates due to different mechanisms from the three planets will help to understand if ion outflow played a role in the early evolution of Mars and Venus as well.

\acknowledgments
The authors acknowledge the support by the Austria Science Fund (FWF) NFN project S116-N16 and the subproject S11604-N16, S11606-N16, and S11607-N16. KK and MG acknowledge the support by the Austrian Research Promotion Agency (FFG) project 873671 ``SmileEarth''. IIA acknowledges the partial financial support by the grant \#RFMEFI61619X0119 of the Ministry of High Education and Science of Russian Federation. MLK is grateful also to the grant No. 18-12-00080 of the Russian Science Foundation and acknowledges the project ``Study of stars with exoplanets'' within the grant No.075-15-2019-1875 from the government of Russian Federation. The software used in this work was partly developed by the DOE NNSA-ASC OASCR Flash Center at the University of Chicago. This research was conducted using resources provided by the Swedish National Infrastructure for Computing (SNIC) at the High Performance Computing Center North (HPC2N), Ume$\mathring{\rm a}$ University, Sweden.  The authors thank the Erwin Schr\"odinger Institute (ESI) of the University of Vienna for hosting the meetings of the Thematic Program ``Astrophysical Origins: Pathways from Star Formation to Habitable Planets'' and Europlanet for providing additional support for this program. The data used in the article are available at the repository 4TU.Centre for Research data: \url{doi:10.4121/uuid:c4ee27c5-7044-4dcb-a3ee-3d204e3160da}.


%
%

\bibliography{EarlyEarthPolarOutflow_revision2.bbl}  

%
%
%
%
%

\end{document}